\documentclass[aps,prl,preprint,superscriptaddress]{revtex4-1}
\usepackage{nicefrac}
\usepackage{braket}
\usepackage{amsmath,amssymb} 
\usepackage{graphicx}
\usepackage[utf8]{inputenc}
\usepackage[T1]{fontenc}
\usepackage{textcomp}

\begin{document}

\title{Tunable generation of entangled photons\\
in a nonlinear directional coupler}

\author{Frank Setzpfandt,$^{1,*}$ Alexander S. Solntsev,$^{1,*,**}$ James Titchener,$^1$ Che Wen Wu,$^1$ Chunle Xiong,$^2$ Roland Schiek,$^3$ Thomas Pertsch,$^4$ Dragomir N. Neshev,$^1$ and Andrey A. Sukhorukov$^1$}

\noaffiliation
\affiliation{
Centre for Ultrahigh Bandwidth Devices for Optical Systems (CUDOS) and Nonlinear Physics Centre,
Research School of Physics and Engineering, Australian National University,
Canberra 0200, Australian Capital Territory, Australia\\
$^2$CUDOS, the Institute of Photonics and Optical Science (IPOS), School of Physics, University of Sydney,
New South Wales 2006, Australia\\
$^3$Ostbayerische Technische Hochschule Regensburg, Pr\"ufeninger Strasse 58, 93049 Regensburg, Germany\\
$^4$Institute of Applied Physics, Abbe Center of Photonics, Friedrich-Schiller-Universit\"at Jena, Max-Wien-Platz 1, 07743 Jena\\
$^{*}$These authors contributed equally\\
$^{**}$Corresponding author: Alexander.Solntsev@anu.edu.au}
\noaffiliation





\begin{abstract}
\noindent The on-chip integration of quantum light sources has enabled the realization of complex quantum photonic circuits. However, for the practical implementation of such circuits in quantum information applications it is crucial to develop sources delivering entangled quantum photon states with on-demand tunability. Here we propose and experimentally demonstrate the concept of a widely tunable quantum light source based on spontaneous parametric down-conversion in a nonlinear directional coupler. We show that spatial photon-pair correlations and entanglement can be reconfigured on-demand by tuning the phase difference between the pump beams and the phase mismatch inside the structure. We demonstrate the generation of split states, robust N00N states, various intermediate regimes and biphoton steering. This fundamental scheme provides an important advance towards the realization of reconfigurable quantum circuitry.
\end{abstract}


\maketitle

\noindent Integrated quantum optics is a promising platform for the precise control of quantum phenomena. In recent years, many integrated components have been developed to manipulate optical quantum states, including beam splitters, phase shifters and quantum logic gates~\cite{Politi:08,Crespi:11,Corrielli:14,Heilmann:14}. By combining such basic elements, the first quantum functionalities and algorithms have been experimentally realized, such as Shor's factoring algorithm~~\cite{Politi:09}, the integrated preparation and measurement of quantum states~\cite{Shadbolt:12}, optical Boson sampling~\cite{Broome:13,Spring:13,Tillmann:13}, and quantum teleportation~\cite{Metcalf:14}.

A key component for integrated quantum optics is the on-chip source of entangled photons. Although different designs of entangled photon sources have been realized~\cite{Tanzilli:12,Clark:14}, they are usually optimized to produce just one predefined output state. However, a true quantum logic requires a flexible quantum architecture, therefore a source able to produce different quantum states on demand~\cite{Valencia:07, Cohen:09}. In particular, quantum photon states with tunable entanglement would be invaluable. Recently, on-chip quantum state sources have been theoretically proposed~\cite{Lugani:12} and the generation of different fully entangled quantum states was experimentally demonstrated~\cite{Silverstone:14,Jin:14}. The tunability in such schemes was achieved by post-processing of the photon pairs generated in uncoupled waveguides by interfering them in separate linear optical couplers with thermo-optically or electro-optically controllable phase shifts.
%

Here we propose a qualitatively new quantum-source concept, where post-processing of the generated photons is supplanted by manipulation of the classical pump beam and generation of the target quantum state in a tunable nonlinear element. Our source is based on photon-pair generation in coupled nonlinear waveguides, which enables on-demand reconfigurable spatial photon-pair correlations and entanglement by tuning the phase difference between the pump beams and the phase mismatch between the pump and the generated photon pairs.

\begin{figure}[t]
\centering
{\includegraphics[width=9.7cm]{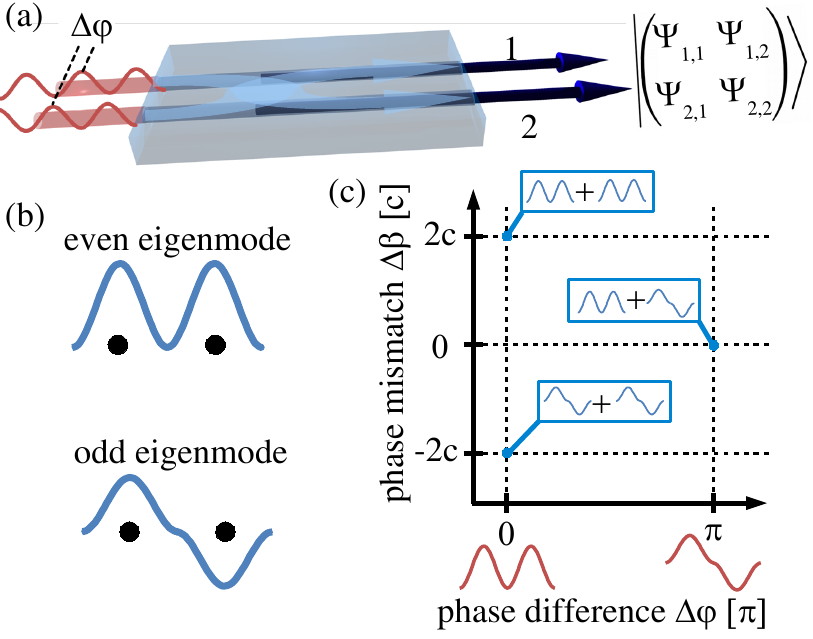}}
\caption{(a) Sketch of the proposed scheme for tunable generation of biphoton quantum states, with two pump beams (red) exciting a directional coupler and generating photon pairs (blue) through SPDC. The output state is described by a wavefunction in the spatial basis. (b) Schematic representation of even (upper panel) and odd (lower panel) eigenmodes of the coupler. (c) Parameter space of the phase mismatch and input phase difference where points of phase-matching to specific combinations of signal and idler eigenmodes are marked by the blue dots.}
\label{fig:intro}
\end{figure}

%
While photon-pair generation in coupled optical waveguides has been already studied theoretically~\cite{Solntsev:12e, Solntsev:12d, Hamilton:14} and experimentally~\cite{Kruse:13, Solntsev:14}, these schemes were mainly considered for the purpose of quantum-walk based simulations. 
Here we demonstrate how the flexibility of coupled waveguide systems can be harnessed for precise control of the process of photon-pair generation, enabling an on-chip biphoton source with unprecedented quantum state tunability. 

To experimentally demonstrate the power of this concept, we use two identical parallel waveguides with evanescent mode coupling as shown in Fig.~\ref{fig:intro}(a), fabricated in LiNbO$_3$, a material with high second-order nonlinearity. Two waveguides are sufficient to generate spatially entangled Bell states (states with maximum entanglement) and demonstrate the proposed wide-range tunability, although the same concept can be extended to a larger number of waveguides. Photon pairs are generated by spontaneous parametric down-conversion (SPDC)~\cite{Burnham:70}, where the properties of the generated pairs depend on the phase mismatch $\Delta\beta$ in the waveguides. The generated signal and idler photons can hop between the waveguides due to evanescent coupling, whereas the pump beams are confined to their excitation sites. Such a system has been theoretically shown to generate N00N states when pumped in one waveguide~\cite{Kruse:14}. Here we consider a general scheme, where the coupler is driven by two classical pump beams, one coupled into each waveguide, with a variable phase difference $\Delta\varphi$ between them. Depending on the phase difference $\Delta\varphi$ and the phase mismatch $\Delta\beta$ various output states are possible~\cite{Wu:14}. Here, the phase mismatch $\Delta \beta$ refers to the single waveguide without the presence of coupling. We note that our approach does not need additional linear optical elements to post-process the generated photons, thus reducing unwanted quantum decoherence. Furthermore, we do not rely on the indistinguishability of photons from different sources. Hence, our concept is easily scalable to higher numbers of waveguides enabling more degrees of freedom. A similar concept for the generation of tunable quantum states with spectral encoding was proposed recently~\cite{Kumar:2014:NatComm}.

Although we target spatially encoded output states, the operation principle of our source is more readily understandable in eigenmode space. The system of two coupled waveguides has two eigenmodes, symmetric (even) and anti-symmetric (odd), which have a phase difference of 0 or $\pi$, respectively, between the field in the two waveguides. These eigenmodes are schematically depicted in Fig.~\ref{fig:intro}(b). By changing the pump phase difference the pump can be continuously tuned between these two modes. SPDC depends on both the overlap of the pump, signal, and idler modes and the phase mismatch $\Delta \beta$, as schematically shown in Fig.~\ref{fig:intro}(c). For an even pump mode with $\Delta \varphi = 0$, a high nonlinear overlap is only possible to even-even or odd-odd combinations of signal and idler modes due to the mode symmetries. However, the evenescent coupling at the respective wavelength leads to a splitting of the propagation constants of these modes. Hence, efficient nonlinear interaction with the two different signal / idler combinations takes place for different phase mismatches, which are determined by the coupling strength $c$ to $\Delta\varphi=\pm 2c$. For odd pump field with $\Delta \varphi = \pi$, efficient interaction is only possible with even-odd combinations of signal and idler modes at a phase mismatch of zero. Since for low powers SPDC depends linearly on the pump, linear combinations of the described states can be achieved within the two-dimensional parameter space defined by $\Delta\beta$ and $\Delta\varphi$. We would like to emphasize that the coupling between waveguides enables this truly two-dimensional parameter space, as the coupling induced mode splitting allows to independently control the probabilities for generation of even-even and odd-odd photon pairs and furthermore enables nontrivial interference between these generated pairs.


For near-degenerate signal and idler photons, the evolution of the biphoton wavefunction in physical space $\ket{\Psi^{\left(\text{n}\right)}}$ is described by the following normalized coupled-mode equations~\cite{Grafe:12}
\begin{eqnarray}
\label{coupledmode}
&i\frac{\partial }{\partial z}\Psi^{\left(\text{n}\right)}_{1,1}&=-c\left[\Psi^{\left(\text{n}\right)}_{1,2}+\Psi^{\left(\text{n}\right)}_{2,1}\right]+i\gamma A_1 \exp\left(i\Delta\beta z\right), \\
&i\frac{\partial }{\partial z}\Psi^{\left(\text{n}\right)}_{2,2}&=-c\left[\Psi^{\left(\text{n}\right)}_{2,1}+\Psi^{\left(\text{n}\right)}_{1,2}\right]+i\gamma A_2 \exp\left(i\Delta\varphi\right) \exp\left(i\Delta\beta z\right), \nonumber \\
&i\frac{\partial }{\partial z}\Psi^{\left(\text{n}\right)}_{2,1}&=i\frac{\partial }{\partial z}\Psi^{\left(\text{n}\right)}_{1,2}=-c\left[\Psi^{\left(\text{n}\right)}_{1,1}+\Psi^{\left(\text{n}\right)}_{2,2}\right]. \nonumber
\end{eqnarray}
Here, 
$\Psi^{\left(\text{n}\right)}_{l,m}$ is the probability amplitude for the signal and idler photons to be in waveguide $l$ and $m$, respectively, $\gamma$ is the generation probability of a photon pair proportional to the effective quadratic nonlinearity and the $A_{\nicefrac{1}{2}}$ are the classical pump amplitudes in waveguides $\nicefrac{1}{2}$. 
To solve these equations we make a transformation
\begin{equation}
\label{transform}
\Psi^{\left(\text{n}\right)}_{l,m}=\sum_{k_l,k_m=\left[0,\pi\right]} \Psi^{\left(\text{k}\right)}_{k_l,k_m}\exp\left(ik_ll\right)\exp\left(ik_mm\right),
\end{equation}
where $\ket{\Psi^{\left(\text{k}\right)}}$ is the wavefunction in eigenmode space. Inserting this into the coupled mode equation Eq.~\ref{coupledmode} allows for the analytic calculation of the wavefunction in eigenmode space, resulting in
\begin{eqnarray}
\label{solution_fourier}
\Psi^{\left(\text{k}\right)}_{\nicefrac{0,0}{\pi,\pi}}=\frac{\gamma L c}{4 c} \left[A_1 + A_2 {\rm e}^{i\Delta\varphi}\right]\exp\left[i L c \frac{\Delta \beta /c \pm 2}{2}\right]  \\
 \times \text{sinc} \left[L c \frac{\Delta\beta / c \mp 2}{2\pi}\right], \nonumber \\
\Psi^{\left(\text{k}\right)}_{\nicefrac{0,\pi}{\pi,0}}=-\frac{\gamma L c}{4 c} \left[A_1 - A_2 {\rm e}^{i\Delta\varphi}\right]\exp\left[i L c \frac{\Delta\beta / c}{2}\right] \\
 \times \text{sinc} \left[L c\frac{\Delta\beta  /c}{2\pi}\right]. \nonumber
\end{eqnarray}
Here $L$ is the sample length and $\text{sinc}\left(x\right)=\sin\left(\pi x\right)/x$. From this solution it is clear that both the amplitude and the phase of the generated state can be controlled by changing $\Delta\beta$ and $\Delta\varphi$. Transformation back to the waveguide basis is achieved using Eq.~\ref{transform}.
\begin{figure}[tb]
\centering
{\includegraphics[width=9.7cm]{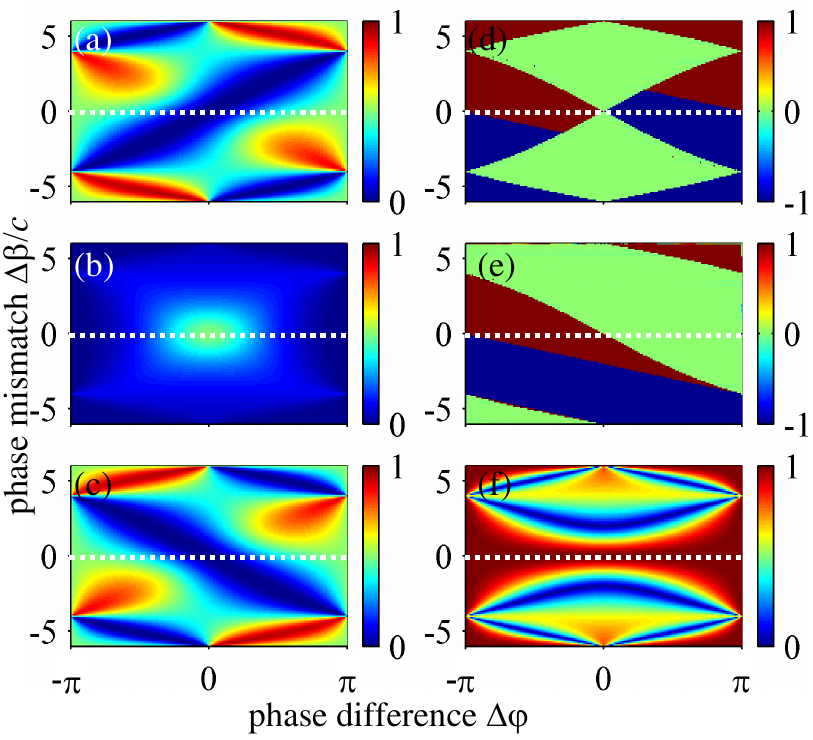}}
\caption{Analytically calculated output quantum state of our source in dependence on the pump phase difference $\Delta\varphi$ and the normalized phase mismatch $\Delta\beta / c$ for $A=1$, $\gamma=1$, and $L c=\pi / 2$. (a-c) Normalized correlation probabilities $\left|\Psi_{1,1}\right|^2$, $\left|\Psi_{1,2}\right|^2=\left|\Psi_{2,1}\right|^2$, and $\left|\Psi_{2,2}\right|^2$, respectively. (d,e) Relative phases of $\Psi_{1,1}$ and $\Psi_{1,2}$ with respect to $\Psi_{2,2}$. (f) Concurrence of the quantum state as a measure for entanglement.}
\label{fig:state}
\end{figure}

In the following we consider in detail a parameter space limited to equal pump intensities, $A_1=A_2=A$ and a device length corresponding to the coupling length, $L c=\pi/2$. As we show below, even with these restrictions a wide-range tunability can be achieved, including the generation of strongly asymmetric states. Calculated wavefunctions, normalized to $\sum_{l,m}\left|\Psi^{\left(n\right)}_{l,m}\right|^2=1$, are shown in Fig.~\ref{fig:state}(a-e).
For the special case of $\Delta\beta=0$, denoted by the white dotted lines in Fig.~\ref{fig:state}, the wavefunction in physical space is
\begin{equation}
\label{state_real}
\Ket{\Psi^{\left(\text{n}\right)}}=-i\frac{\gamma\pi A}{2c} \exp\left(i\frac{\Delta\varphi}{2}\right)
\begin{bmatrix}
\sin\left(\nicefrac{\Delta\varphi}{2}\right) & \hspace{-2mm} \cos\left(\nicefrac{\Delta\varphi}{2}\right) \\
\cos\left(\nicefrac{\Delta\varphi}{2}\right) & \hspace{-2mm} -\sin\left(\nicefrac{\Delta\varphi}{2}\right)
\end{bmatrix}
.
\end{equation}
Hence, our source can be tuned between the Bell states $\ket{\Phi^{\left(+\right)}} \sim
\left(\begin{smallmatrix}
0 & 1 \\
1 & 0
\end{smallmatrix}\right)$ and $\ket{\Psi^{\left(-\right)}} \sim
\left(\begin{smallmatrix}
1 & 0 \\
0 & -1
\end{smallmatrix}\right)$
by changing the pump phase difference from $\Delta\varphi=0$ to $\Delta\varphi=\pi$. These and all other states described by Eq.~\ref{state_real} are maximally entangled states, which can also be generated with other recently demonstrated integrated photon-pair sources with complex post-processing~\cite{Silverstone:14,Jin:14}. 
However, the interference between the process of photon-pair generation and the coupling dynamics between the waveguides used in our work allows for much wider tunability of the output state. Importantly, our source enables control of entanglement of the generated state. This is a prerequisite for the systematic generation of mixed states~\cite{Shadbolt:12}, which is an important resource for certain quantum computation schemes~\cite{Lanyon:2008:PhysRevLett} and for the investigation of transport phenomena~\cite{Plenio:2008:NewJouPhys}. To verify entanglement control we calculated the concurrence \cite{Wootters:1998:PRL}, which is plotted in Fig.~\ref{fig:state}(f). A concurrence of 1 represents the maximal possible entanglement. As expected we find this for all states with $\Delta\beta=0$. Additionally, all states with $\Delta\varphi=-\pi,\pi$, for which $\Psi_{\nicefrac{1,2}{2,1}}=0$, have concurrence of 1 and are proportional to the maximally entangled state $\ket{\Psi^{\left(-\right)}}$. States with smaller entanglement can be generated for all other input phase differences by introducing a phase mismatch. For example, factorisable states with equal detection probability in each waveguide or states with all photons in one waveguide are possible for $\Delta\varphi=0$, $\Delta\beta/ c=\pm2$ and $\Delta\varphi=\pm0.53 \pi$, $\Delta\beta/ c=\pm5$, respectively. In particular the latter states may be useful to steer generated photon pairs in integrated optical circuits.

To experimentally realize the proposed source we use a two-waveguide coupler in lithium niobate~\cite{Schiek:98}. Two individually adjustable pump beams of 671~nm wavelength with tunable relative phase were generated using a Sagnac polarization interferometer (see supplementary material section 1 for details) and coupled to the sample with a microscope objective. Type~I phasematching between TE (pump) and TM (signal/idler) modes is achieved for a sample temperature of around 370~\textdegree C. By changing the temperature we controlled the phase mismatch $\Delta\beta$, where a temperature change of 0.1 \textdegree C corresponds to a change of the phase mismatch of about $2.5c$. We note that $\Delta\beta$ can be also controlled by changing the pump wavelength~\cite{Solntsev:14}. The length of the coupler of $L=47.5\;\text{mm}$ was matched to the signal/idler coupling constant $c=33\;\text{m}^{-1}$ to achieve $L c =\pi/2$ as in the theoretical analysis above. The broad spectral distribution of photons generated by SPDC~\cite{Solntsev:14} is filtered by a bandpass with 12~nm bandwidth to achieve nearly degenerate photon pairs.

\begin{figure}[tb]
\centering
{\includegraphics[width=9.7cm]{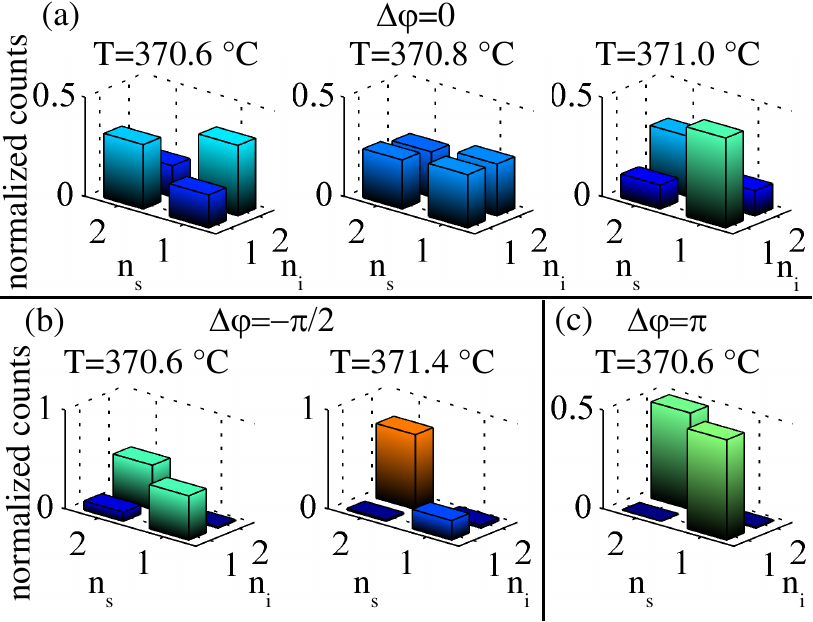}}
\caption{Normalized biphoton correlations for different output states. (a) Tunable quantum state generation for $\Delta\varphi=0$ and different phase mismatches $\Delta \beta$, showing the tunability from anti-bunching (left) to bunching (right). (b) Steering of photon pairs for $\Delta\varphi= -\pi/2$, from even distribution between output waveguides to predominant emission from waveguide 2. (c) Generation of bunched N00N-state for $\Delta\varphi=\pi$.}
\label{fig:correlation}
\end{figure}

The output of our photon-pair source is characterized by correlation measurements over the two possible spatial positions of the photons and by quantum state tomography~\cite{James:01}. Fig.~\ref{fig:correlation} shows selected correlation measurement results, which illustrate the tunability of our source. In Fig.~\ref{fig:correlation}(a) we demonstrate output state tuning by changing the sample temperature 
for pump beams of equal phase. For a temperature of 370.6~\textdegree C, corresponding to a phase mismatch of $\Delta\beta=0$, strong anti-bunching is observed, resembling the Bell-state $\ket{\Phi^{\left(+\right)}}$. With growing temperature and phase mismatch, an increasing amount of bunching is measured, leading to a state with equal count numbers for all waveguide combinations and finally to dominant bunching. These results are in good qualitative agreement with our analytic theory, however, both bunching and anti-bunching are not complete and the wavefunction is broadened with respect to the mismatch. This is due to sample inhomogeneities and the influence of the finite bandwidth filter used in the experiments (see supplementary material section 2 for details), which inhibit the complete destructive interference of the respective smaller components of the wavefunction.

Steering of the generated photon pairs into just one waveguide occurs when either the $\Psi_{1,1}$ or $\Psi_{2,2}$ component of the wavefunction is suppressed and there is negligible contribution from the off-diagonal components. 
Our analytic solution predicts that the steering can be controlled by either the phase mismatch $\Delta\beta$ or the phase difference between the pump beams $\Delta\varphi$.
In the experiment we demonstrate photon-pair steering for a pump phase difference of $\Delta\varphi= -\pi/2$ in Fig.~\ref{fig:correlation}(b). At a temperature of 370.6~\textdegree C, signal and idler photons are leaving the structure together either from the first or from the second waveguide. When the temperature is changed to 371.4~\textdegree C, most photon pairs are detected in waveguide~2, even though both waveguides are pumped with the same intensity.

\begin{figure}[tb]
\centering
{\includegraphics[width=9.7cm]{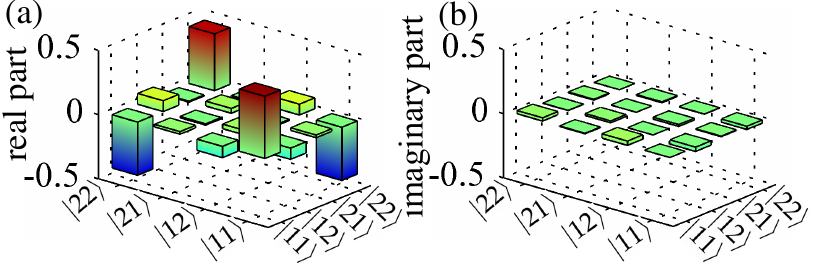}}
\caption{(a) Real and (b) imaginary parts of the measured density matrix of a N00N-state for $\Delta\varphi=\pi$ and $\Delta\beta=0$.}
\label{fig:tomography}
\end{figure}

Finally, for $\Delta\varphi=\pm\pi$ we find complete bunching of the generated photons depicted in Fig.~\ref{fig:correlation}(c). This output state corresponds to a 2-photon N00N-state, similar to the theoretical result reported in~\cite{Kruse:14}. However, in our case $\left|\Psi_{1,1}\right|=\left|\Psi_{2,2}\right|$ and $\left|\Psi_{1,2}\right|=\left|\Psi_{2,1}\right|=0$ for all mismatches since the excited odd pump mode always results in even-odd combinations of signal and idler modes, i.e. anti-bunching in eigenmode space, leading to spatial bunching. We experimentally confirm this robustness for varying sample temperatures (see supplementary material section 3 for details). To verify that the generated state corresponds to $\ket{\Psi^{\left(-\right)}}$ we performed quantum tomography. Amplitude and phase of the measured density matrix are plotted in Fig.~\ref{fig:tomography}. The experimental fidelity with the ideal N00N state is $0.93 \pm 0.03$ (see supplementary material section 3 for details), and the concurrence is $0.79 \pm 0.08$.

In summary, we analyzed theoretically and demonstrated experimentally a new concept for a flexible biphoton quantum state source. Photon pair generation in this source is based on SPDC in a nonlinear directional coupler, which is pumped by two independent pump beams. The evanescent coupling of the generated photons allows for a much larger tunability of the generated quantum states compared to previous realisations using uncoupled waveguides as photon sources~\cite{Silverstone:14,Jin:14}. The phase difference between these pumps and the SPDC phase mismatch provide two control parameters which are used to determine the produced quantum state. No post-processing is required. We show the flexibility of our approach by experimentally demonstrating the generation of anti-bunched and bunched Bell-states, photon pair steering to one output, and the generation of maximally entangled N00N states. We note that the phase mismatch, which we controlled by changing the sample temperature, could also be tuned by changing the pump wavelength~\cite{Solntsev:14}. In this case, all-optical control of the output quantum state can be achieved by only altering the properties of the pump light. Furthermore, although we use a bulk optical setup to prepare the two pump beams with variable phase difference, this functionality could also be integrated on a chip using electro-optic phase shifters~\cite{Jin:14}. Thus, the pump phase could be quickly modulated, enabling e.g. photon pair switching with telecom signal speeds. We believe that the demonstrated concept can be an important building block for integrated quantum optics, increasing the flexibility in tailoring the output state.

This work was supported by the Australian Research Council (DP130100135, FT100100160, DE12010022 and CE110001018), the German Research Foundation (PE 1524/5-2), and the German Academic Exchange Service (56265492).


\section{Supplementary material}

\subsection{1. Experimental setup}
\begin{figure}[b]
\centering
{\includegraphics[width=9.7cm]{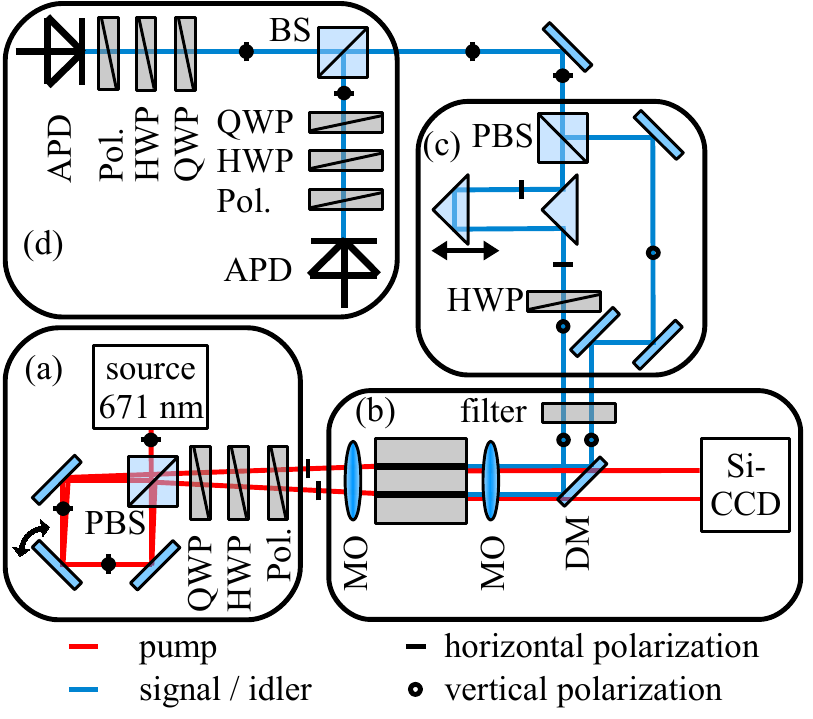}}
\caption{Scheme of the used setup with (a) pump beam preparation, (b) quantum state generation, (c) transfer from spatial basis to polarization basis, and (d) quantum tomography. Used abbreviations are: PBS - 50:50 polarization beamsplitter, BS - 50:50 non-polarizing beamsplitter, QWP - quarter-wave plate, HWP - half-wave plate, Pol. - polarizer, MO - microscope objective, DM - dichroic mirror, filter - spectral filter, and APD - avalanche photo diode.}
\label{fig:setup}
\end{figure}

The setup used for two-photon state generation and analysis is schematically depicted in Fig.~\ref{fig:setup}.  Pump light is provided by a continuous wave laser with 671~nm wavelength, which polarization is diagonal, i.e. has equal intensity in the horizontal and vertical polarization components. For the independent preparation of two classical pump fields with tunable relative phase we utilize a Sagnac polarization interferometer as outlined in Fig.~\ref{fig:setup}(a). One of the mirrors can be tilted to allow for spatial alignment of the orthogonally polarized beams propagating clock- and counterclockwise through the interferometer. The relative phase of the two spatially separated beams is controlled by a quarter-wave plate and a half-wave plate after recombination of the clock- and counterclockwise propagating beams. Finally, a polarizer selects only the horizontal polarization components, allowing for efficient excitation of the TE pump modes in the coupler. This pump preparation setup allows for a very high stability of the phase difference between the two beams since they travel almost the same optical paths in the interferometer. No additional stabilization of the interferometer was used.

To generate the photon pairs, the two pump beams are coupled to the sample by one microscope objective as shown in Fig.~\ref{fig:setup}(b). To fulfill the phase-matching conditions, the sample was heated to approximately 371~\textdegree C using an oven with three independent heating elements controlled by a high precision temperature controller (Stanford Research Systems PTC10). To control the phase mismatch, the temperature in the oven was changed. A temperature difference of 0.1 \textdegree C corresponds to a change of the phase mismatch of about 85/m or $2.5c$, with $c=33/\text{m}$ being the coupling constant of the signal and idler modes. At the end of the sample, signal and idler photons and the remaining pump were extracted by another microscope objective and separated by a dichroic mirror. The pump coupling was continuously monitored with a silicon camera (Philips SPC900NC). Furthermore, an interference filter with FWHM of 12~nm (ThorLabs FB1340-12) was used to filter the photon pairs around the degeneracy wavelength of 1342~nm.

To analyze the quantum states we perform quantum tomography in the polarization basis \cite{James:01}. To this end we transform the state, which is generated in the spatial basis of the two waveguides, into the polarization basis with the interferometer depicted in Fig.~\ref{fig:setup}(c). In the image plane of the out-coupling microscope objective the photons from the two waveguides, which are both TM polarized, are separated by a D-shaped mirror. The polarization of one arm is rotated to TE and the two beams are recombined with a polarizing beam splitter. To compensate any path difference between the two polarization components we additionally installed an adjustable delay line. Quantum tomography is then performed by splitting the beam again using a 50:50 beamsplitter as shown in Fig.~\ref{fig:setup}(d). In each arm a polarization analyzer is placed, after which the photons are coupled to single mode fibers and routed to single photon detectors (IDQuantique ID210) based on InGaAs avalanche photo diodes. These are connected to a time-to-digital converter (RoentDek HM-1) to measure the two-photon coincidences with a timing resolution of 133~ps. The same setup was used to measure correlations only, which corresponds to a measurement with the polarization analyzers in the four possible combinations of horizontal and vertical polarization.

\subsection{2. Sample inhomogeneity and spectral filter}

\begin{figure}[b]
\centering
{\includegraphics[width=9.7cm]{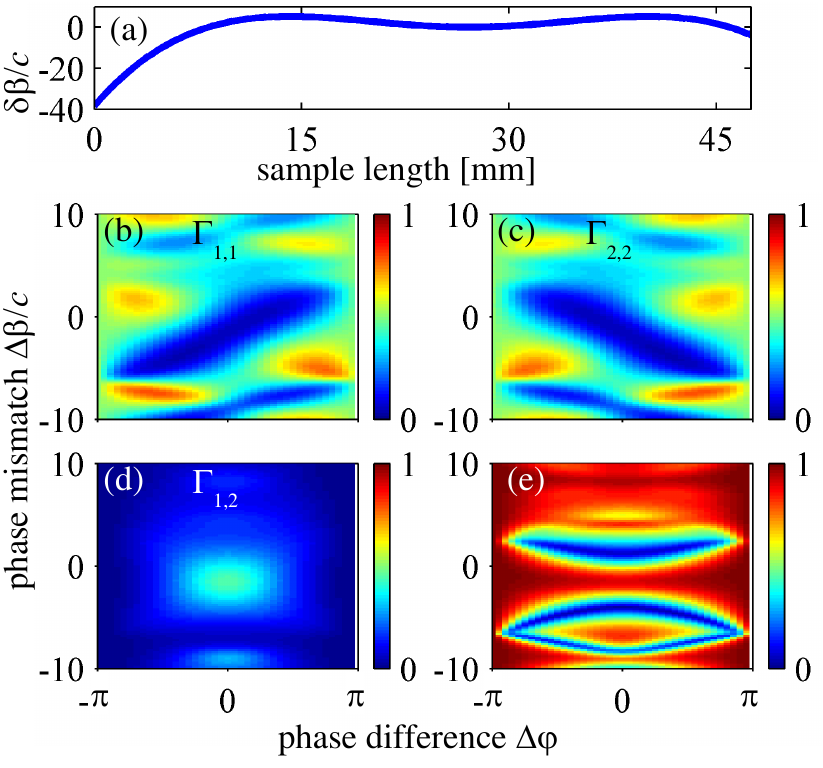}}
\caption{(a) Normalized local phase mismatch difference $\delta\beta / c$ in the heated sample as determined by SHG measurements in a single waveguide. (b-e) Simulated output quantum state in inhomogeneous sample with finite filter bandwidth in dependence on the pump phase difference $\Delta\phi$ and the normalized phase mismatch $\Delta\beta / c$ for $A = 1$, $\gamma = 1$, and $L c = \pi/2$. (b-d) Normalized correlations $\left|\Gamma_{1,1}\right|$ ,$\left|\Gamma_{2,2}\right|$ and $\left|\Gamma_{1,2}\right| = \left|\Gamma_{2,1}\right|$ , respectively, and (e) Corresponding concurrence.}
\label{fig:inhom}
\end{figure}

Although the measured coincidences described in the main part of this paper agree qualitatively with the analytically predicted quantum state, there exist some notable differences. For example, we could not achieve complete anti-bunching. This is due to technical details of the used sample and setup, the influence of which we investigate in the following part of the supplementary material. In particular, we discuss the influence of inhomogeneities introduced by the sample oven and the effect of the spectral filter used to select near-degenerate signal and idler photons.

To achieve phase matching the temperature of the sample had to be elevated to about 371 \textdegree C. To this end, the sample is stored in an isolated oven, which, however, is open at the end facets to enable efficient light coupling. This leads to a temperature drop at the sample edges. Hence, there is an additional phase mismatch $\delta\beta\left(z\right)$, corresponding to the position-dependent temperature inhomogeneity. The temperature inhomogeneity is characterized by measuring the second-harmonic generation (SHG) in dependence on the average sample temperature. The width of the obtained SHG tuning curve is a measure of the inhomogeneity~\cite{Schiek:98}. By introducing different setpoints for the three individual heating regions of our oven we minimized the width of the tuning curve. The optimal setting had temperature setpoints raised by 0.6~\textdegree C in the outer heating regions with respect to the center of the oven. The actual temperature profile was estimated by carefully matching the measured tuning curve with the simulations that took the temperature inhomogeneity into account using a polynomial model \cite{Schiek:98}. The resulting inhomogeneity $\delta\beta\left(z\right)$ is shown in Fig.~\ref{fig:inhom}(a). The largest deviation of this curve from the value in the center of the sample corresponds to a temperature drop of about 1.5~\textdegree C. The asymmetry of the inhomogeneity profile is caused by a shift of the sample towards the in-coupling objective, which was necessary to obtain optimal coupling of the pump beam through the thick insulation layers of the oven.

The analytic calculations in the main part of this paper are valid only for almost exactly degenerate photon pairs. In general, the photon pair states produced in our coupler are spectrally broad. To work in a near-degenerate regime, we placed an interference filter with a FWHM bandwidth of 12~nm after the oven. Within this bandwidth, all coefficients of the coupled mode equations (Eq.~1 of the main document) are constant, with exception of the phase mismatch, which due to narrow phase matching bandwidth is very sensitive to small changes in wavelengths. The total phase mismatch can be written as follows:
\begin{equation}
\Delta\beta\left(\omega\right)=\Delta\beta_0+\Delta\beta_\omega\left(\omega\right)+\delta\beta\left(z\right),
\end{equation}
where $\Delta\beta_0$ is the mismatch for degenerate SPDC defined by the overall sample temperature (the value used in the main document), $\delta\beta\left(z\right)$ is induced by the temperature inhomogeneity along the sample, and $\Delta\beta_\omega\left(\omega\right)$ is the additional phase mismatch induced by the non-degenerate signal and idler frequencies according to the experimentally measured dispersion for this sample~\cite{Schiek:98}. Here $\omega = \omega_s$ is the signal frequency, which also defines the idler frequency $\omega_i = \omega_p-\omega$ though the monochromatic pump frequency $\omega_p$. 


The resulting set of the coupled mode equations uses a modified phase mismatch $\Delta\beta$, and a biphoton wavefunction $\Psi(\omega)$ depending on the frequency $\omega$ as follows:
\begin{eqnarray}
\label{coupledmode_inhom}
i\frac{\partial }{\partial z}\Psi^{\left(\text{n}\right)}_{2,1}\left(\omega\right)=i\frac{\partial }{\partial z}\Psi^{\left(\text{n}\right)}_{1,2}\left(\omega\right)=-c\left[\Psi^{\left(\text{n}\right)}_{1,1}\left(\omega\right)+\Psi^{\left(\text{n}\right)}_{2,2}\left(\omega\right)\right]. \nonumber \\
i\frac{\partial }{\partial z}\Psi^{\left(\text{n}\right)}_{1,1}\left(\omega\right)=-c\left[\Psi^{\left(\text{n}\right)}_{1,2}\left(\omega\right)+\Psi^{\left(\text{n}\right)}_{2,1}\left(\omega\right)\right] \nonumber \\
+i\gamma A_1 \exp[i\int_{0}^{z}\Delta\beta(\omega,z')dz'], \\
i\frac{\partial }{\partial z}\Psi^{\left(\text{n}\right)}_{2,2}\left(\omega\right)=-c\left[\Psi^{\left(\text{n}\right)}_{2,1}\left(\omega\right)+\Psi^{\left(\text{n}\right)}_{1,2}\left(\omega\right)\right] \nonumber \\
+i\gamma A_2 \exp[i\int_{0}^{z}\Delta\beta(\omega,z')dz'], \nonumber 
\end{eqnarray}
The correlations taking into account a Gaussian spectral filter with a frequency response $F(\omega)=\exp [-4\ln\left(2\right)\left(\omega-\omega_0\right)^2 / \sigma_{\rm FWHM}^2 ]$, central frequency $\omega_0 = 1342$ nm and bandwidth $\sigma_{\rm FWHM}  = 12$ nm, are calculated according to the following expression:  
\begin{equation}
   \Gamma^{(F)}_{n_s, n_i} = \int d{\omega} F(\omega) F(\omega_p-\omega) |\Psi_{n_s, n_i}^{\left(n\right)}(L, \omega)|^2.
\end{equation}

The correlations of the quantum state produced by the inhomogeneous sample after the spectral fitler are depicted in Fig.~\ref{fig:inhom}(b-d). Due to the inhomogeneity the generated maxima are widened and shifted towards smaller mismatches. Inhomogeneous background is visible for larger mismatches. Although these new features change the correlations, they are still similar to the analytic solution that we investigated in the main paper. In particular, for $\Delta\varphi=\pm\pi$ complete bunching and the generation of N00N states with $\Gamma_{1,2}=\Gamma_{2,1}=0$ is still possible, as evidenced by our experimental results. However, the generation of anti-bunched states in the analytical model relies on destructive interference leading to suppression of the $\Gamma_{1,1}$ and $\Gamma_{2,2}$ components. This interference is inhibited in the experiment, due to both the inhomogeneity and the finite spectral width. In agreement with our experimental results this perturbation of the interference leads to incomplete anti-bunching.

\begin{figure}[b]
\centering
{\includegraphics[width=9.7cm]{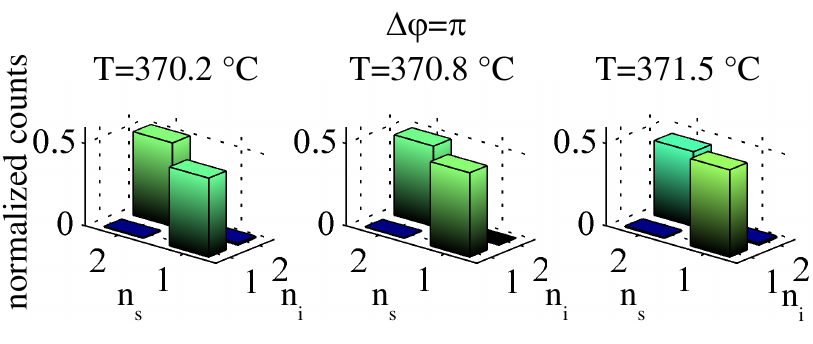}}
\caption{Experimentally measured spatial correlations for the N00N state pumped by the odd pump with $\Delta\phi=\pi$ at different sample temperatures.}
\label{fig:bunching}
\end{figure}

The two-dimensional density matrix of coupler output before filter $\Psi^{\left(n\right)}$ with index $n$ representing all possible combinations of indexes $n_s$ and $n_i$ is calculated as follows:
\begin{equation}
   \rho=\Ket{\Psi^{\left(n\right)}}\Bra{\Psi^{\left(n\right)}} = \Psi^{\left(n\right)} \left[\left(\Psi^{\left(n\right)}\right)^\top\right]^*
\end{equation}
To obtain the two-dimensional density matrix after the filter, each element is multiplied by the full density matrix with the corresponding filter transmission:
\begin{equation}
	\rho_{n_s,n_i,\omega}^{(F)} = F(\omega) F(\omega_p-\omega)  \rho_{n_s,n_i,\omega}
\end{equation}
The reduced density matrix, which describes the state measured by our wavelength insensitive detectors is calculated by tracing over the frequencies:
\begin{equation}
	\rho_{n_s,n_i}^{(reduced)} = \text{Tr}_\omega\left[\rho_{n_s,n_i,\omega}^{(F)}\right]
\end{equation}
Since the filter functions are not normalized, the resulting density matrix is re-normalized as follows:
\begin{equation}
\sum_{n_s,n_i}\rho_{n_s,n_i}^{(reduced)}\left(\rho_{n_s,n_i}^{(reduced)}\right)^*=1
\end{equation}
The concurrence is calculated from the reduced density matrix, following the Ref.~\cite{Wootters:1998:PRL}.
%
%
The possibility of highly entangled state generation is indicated by the concurrence numbers approaching unity, as shown in Fig.~\ref{fig:inhom}(e). 

These simulations of the photon-pair source with inhomogeneity, taking into account the finite filter bandwidth, are in good agreement with our experimental results, explaining the differences from the analytic prediction. We point out that these experimental shortcomings are of purely technical nature and by no means a limitation of our source concept. The source performance could be improved by using a different oven design with better maintained temperature homogeneity, a shorter sample with stronger coupling that is more robust against temperature inhomogeneity, or a narrower spectral filter.

\subsection{3. N00N state}

\begin{figure}[tb]
\centering
{\includegraphics[width=9.7cm]{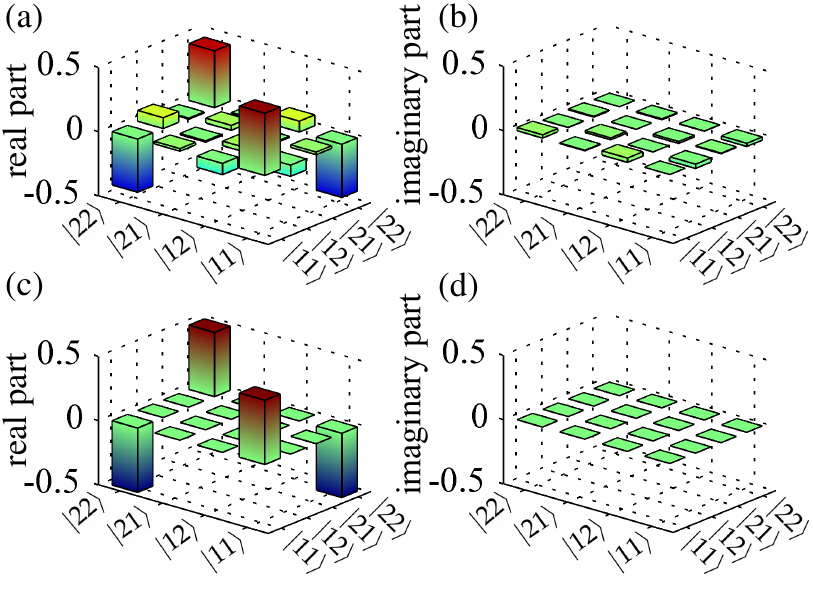}}
\caption{(a,c) Real and (b,d) imaginary parts of the N00N-state  density matrix for $\Delta\varphi=\pi$ measured experimentally (a,b) and calculated analytically (c,d).}
\label{fig:tomography_full}
\end{figure}

The biphoton N00N state generated in a nonlinear directional coupler does not require precise phase matching, and is therefore robust in respect to moderate fabrication inaccuracies and temperature changes. 
In particular, the inhomogeneous temperature distribution along the sample and the presence of spectral filter only affect the generation efficiency, and not the spatial correlations. Therefore the normalized analytic solutions from the main paper for the N00N state are not distinguishable from the simulation results discussed in the previous section. The experimentally measured spatial correlations for the N00N state are shown in Fig.~\ref{fig:bunching} for different phase-matching conditions. It is evident that even relatively large temperature changes do not affect the correlations. The tolerance of the N00N state to perturbations is due to the fact that for an input phase mismatch of $\Delta\varphi = \pm \pi$ only the odd pump eigenmode of the coupler is excited, which enables SPDC only to the even-odd pair of signal and idler eigenmodes. The ensuing anti-bunching in eigenmode space leads to the spatial bunching we observe. This effect does not rely on any interference effects or specific phasematching conditions, hence it is not disturbed by inhomogeneities along the waveguide. However, the quality of the N00N states would decrease if differences between the two waveguides in the coupler break the symmetry of the two-waveguide system.

The comparison of the experimentally measured density matrix for the N00N state [Fig.~\ref{fig:tomography_full}(a,b)] and the corresponding numerical prediction [Fig.~\ref{fig:tomography_full}(c,d)] shows good agreement with fidelity of $0.93 \pm 0.03$ and additionally confirms the robustness of the N00N state generation in a nonlinear directional coupler.

\end{document}